\documentclass[review,12pt]{article}
\usepackage{graphicx}
\usepackage{url}
\usepackage{multirow}
\usepackage{natbib}
\usepackage[table,xcdraw]{xcolor}
\usepackage{amsmath}
\usepackage{array}
\usepackage{soul}
\usepackage{orcidlink}
\usepackage{authblk}

\begin{document}


\title{Location and audience selection for maximizing social influence} 
\date{}
\author[1,2]{Bal\'azs R. Sziklai \texorpdfstring{%
\orcidlink{0000-0002-0068-8920}
}{}}

\author[1,3]{Bal\'azs~Lengyel\texorpdfstring{%
\orcidlink{0000-0001-5196-5599}
}{}}
\affil[1]{Centre for Economic and Regional Studies, Budapest, Hungary}
\affil[2]{Department of Operations Research and Actuarial Sciences, Corvinus University of Budapest}
\affil[3]{Centre for Advanced Studies, Corvinus University of Budapest, Hungary}

\maketitle

\begin{abstract}
Viral marketing campaigns target primarily those individuals who are central in social networks and hence have social influence. Marketing events, however, may attract diverse audience.
Despite the importance of event marketing, the influence of heterogeneous target groups is not well understood yet.
In this paper, we define the Spreading Potential (SP) problem in which different sets of agents need to evaluated and compared based on their social influence. A typical application of SP is choosing locations for a series of marketing events.
The SP problem is different from the well-known Influence Maximization (IM) problem in two aspects. Firstly, it deals with sets rather than nodes. Secondly, the sets are diverse, composed by a mixture of influential and ordinary agents. Thus, SP needs to assess the contribution of ordinary agents too, while IM only aims to find top spreaders.
We provide a systemic test for ranking influence measures in the SP problem based on node sampling and on a novel statistical method, the Sum of Ranking Differences. Using a Linear Threshold diffusion model on an online social network, we evaluate seven network measures of social influence. We demonstrate that the statistical assessment of these influence measures is remarkably different in the SP problem, when low-ranked individuals are present, from the IM problem, when we focus on the algorithm's top choices exclusively.
\end{abstract}

\section{Introduction}
How should we distribute a limited number of product samples or targeted messages in a marketing campaign to achieve the greatest spread on the market by social influence?


Similar questions have been in the spotlight of research on network diffusion in the past two decades. In their seminal paper, \citet{Kempe2003} formulate the \emph{influence maximization problem} (IM) to answer this question as follows: Given a diffusion model $D$, a network $G$ and a positive integer $k$, find the most influential $k$ nodes in $G$ whose activation results in the largest spread in the network. Influence maximization has been studied in various contexts and targeting individuals has been further specialized for real world applications considering seed diversity \citep{CALIO2021}, characteristics of the propagated information \citep{Li2020}, user susceptibility \citep{Xia2021}, evolution of a dynamic network \citep{Greenan2015} and the balancedness of the spread among different attribute groups \citep{Lin2020}. However, it is less understood how the IM method can be applied in traditional marketing techniques, like targeting groups.

Marketing campaigns typically include group targeting in the form of commercials or public events but the message of these potentially spreads further in the social networks of the initial audience \citep{shi2019location, li2011labeled, yan2017group,Saleem2017}. Consider the example that an NGO would like to launch a drug prevention campaign in schools or a politician competes for votes in towns. Due to limited resources, only few high-schools can be visited in the prevention campaign or few speeches can be held before elections. Thus, selecting a handful set of audiences but aiming for a wide reach through social networks, it is important to target groups that have the greatest spreading potential.

Motivated by the frequent combination of target group selection and viral marketing, which is difficult to address in the original IM setting,
we formulate the \emph{Spreading Potential} (SP) problem. Given a diffusion model $D$, a network $G=(V,E)$, assess the collection of node sets $S_i \subset V,\ i=1,\dots,n$ according to the spread that their activation generates in order to find $S_{\hat{i}}$ of maximum spread. We pose no restriction on the subsets, they might be of varying sizes and some might even overlap.

Unlike in the IM problem that focuses on finding the most influential individuals, approaches in the SP problem must handle the challenge that $S_i$ can mix highly influential and ordinary individuals. Hence, a major difference between the IM and SP approaches comes from the influence of ordinary agents in the diffusion model. In Section~\ref{sec:SPvsIM} we discuss the differences of the two models in detail.

Influence in networks can be proxied by the centrality of the individuals (see Section~\ref{sec:IF_proxies}), but certain centrality indices might be better in ranking ordinary agents than others. Put it differently, an influence measure that is successful in identifying the best spreaders on the whole node set might be less efficient in classifying agents from the middle.

In this paper, we devise a statistical test to uncover the real ranking of influence maximization proxies under the SP model. The situation is somewhat similar to polling. When a survey agency wants to predict the outcome of an election it takes a random, representative sample from the population.  Our aim is to observe the performance of the measures on 'average' nodes, thus we will take samples from the node set, then compare the obtained results using a novel comparative test method: Sum of Ranking Differences (SRD or H\'eberger-test).

The SP approach offers various practical advantages. First, it can handle the combination of group targeting and viral marketing, \emph{e.g.}\ when campaigns are built around events or locations (e.g.\ festivals, schools) rather than individuals. Second, the SP could mitigate the problem if the most influential agents do not help the campaign because they are risk-averse unwilling to try the product\footnote{In his classic book, \cite{Rogers2003} classifies adopters into five groups: innovators, early adopters, early majority, late majority and laggards. Half of the population consists of agents from the last two group.}, or even belong to a group that actively opposes the product for ideological reasons (e.g.\ anti-vaxxers, religious groups, groups affiliated with a political party), or when the company does not have access to these agents (they didn't provide contact or give consent to be included).

The proposed method to assess influence proxies under the SP model is SRD, that ranks competing solutions based on a reference point. It originates from chemistry where various properties of a substance measured in different laboratories (or by various methods) need to be compared \citep{Heberger2010,Heberger2011}. SRD is rapidly gaining popularity in various fields of applied science such as analytical chemistry \citep{Andric2018}, pharmacology \citep{Ristovski2018}, decision making \citep{Lourenco2018}, machine learning~\citep{Moorthy2017}, political science \citep{Sziklai2020} and even sports~\citep{West2018}.

Our test method is devised as follows. Let us suppose that the diffusion model $D$ and the network $G$ is fixed. We take samples from the node set of $G$. We run the diffusion model with these samples as input and observe their performance, the latter will serve as a reference ranking for SRD. Then, we calculate the influence maximization proxies that we would like to compare. We aggregate the proxy values for each sample set. This induces a ranking among the samples: different proxies will prefer different sample sets. Finally, we compare the ranking of the sample sets made by the proxies with the reference ranking. The measured distances (SRD scores) show which proxies are the closest to the reference, that is, the most effective.

For demonstration, we use a real-life social network. iWiW was the most widely used social network in Hungary before the era of Facebook \citep{Lengyel2020}. Over its life-cycle, it had more than 3 million subscribers (nearly one-third of the population of Hungary) who engaged in over 300 million friendship ties. For computational reasons, we use a 10\% sample of the node set of this network chosen uniform at random but stratified by towns and network community structure. We compute various influence maximization measures and calculate their ranking under the Linear Threshold diffusion model.

We only know of one paper that uses statistical testing to compare centrality measures. \cite{Perra2008} analyse the rankings of nodes of real graphs for different algorithms. They focus on spectral measures such as PageRank and HITS and adopt Kendall's $\tau$ index to calculate their pairwise correlation. They do not use any external validation, hence they do not derive any ranking from their analysis. Theoretical differences, taxonomy and computational complexity of IM algorithms are discussed in the survey paper of \cite{Li2018}. A detailed introduction on the Influence Maximization Problem and network diffusion models can be found in \cite{Chen2013}.

\section{Spreading potential vs Influence maximization}\label{sec:SPvsIM}

\begin{figure}
    \centering
    \includegraphics[width=6cm]{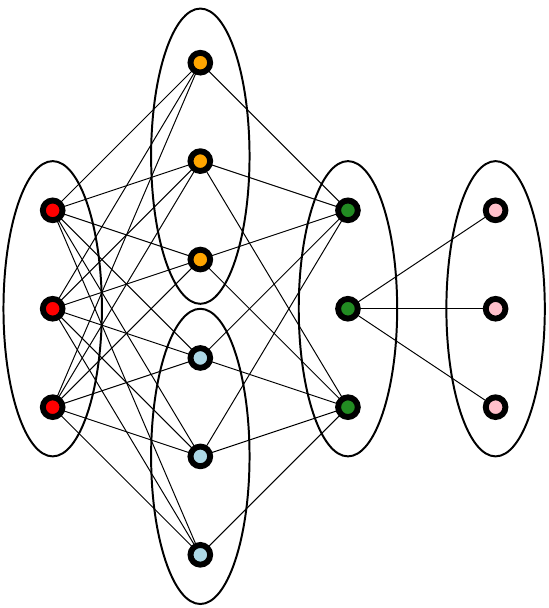}
    \caption{Sketch of a social network with five geographical locations (red, orange, blue, green and pink) each accommodating  three agents.}
    \label{fig:towns}
\end{figure}

Since the proposed problem is closely related to influence maximization, it is worth to explore the differences between the two approaches. Let us consider a small example. Figure~\ref{fig:towns} depicts a social network with additional geographic information. Colors signify different locations, say towns. Each town has three inhabitants (agents). Red agents are all connected to all the orange and blue agents. The top green agent is connected to all the orange agents and two blue ones, while the bottom green is connected to all the blue agents and two orange. Finally, the middle green agent is connected to all the pink agents.

In the spreading potential problem we have to find the most suitable locations for a series of marketing events. Agents corresponding to the same location can be reached simultaneously. Let us assume that the linear threshold diffusion model (a standard choice in the literature, see details in Section~\ref{sec:diff_models}) adequately represents influence spreading in this network. In our example one event is organised. Which town should be chosen?

Formulating the question like this, we are asking about the spreading potential of the towns. In contrast, the influence maximization problem tries to find the $k$ most pivotal agents in a diffusion.

Let us try to answer our questions from the perspective of influence maximization. Since each town accommodates three agents, let us fix $k=3$. \cite{Kempe2003} proposed a greedy algorithm to approximate the best set of $k$ agents. Since then many clever heuristics has been invented to improve either the approximation or the running time of the greedy algorithm. However, for such a small example we won't need sophisticated techniques. We can try each combination of agent-triplets in a diffusion simulation. It turns out, that choosing any two of the red agents, and the middle green agent is the best: On average, they can reach (activate) 78.6\% of the nodes.

Since the majority of the most influential agents are red, we are inclined to choose the red town. However, that would be a mistake. Green agents can activate on average 74.0\% of all the nodes, while red agents only score 63.6\%. Thus, the green town outperforms the red one by a hefty 10\%.

Let us approach the question from another angle. Influence maximization techniques often suffer from computational limitations. It is difficult to even approximate a suitable $k$ set for a huge network. A usual workaround is to use proxy-measures, that is, network centralities that were designed to find influential spreaders. For instance, we may calculate the average PageRank value of each agent in a town (for definition and discussion of network centralities, see Section~\ref{sec:IF_proxies}). Table~\ref{tab:towns} contains the PageRank values as well as the Harmonic centrality of each node. It also lists the average centrality values for each town.

\begin{table}
\caption{\label{tab:towns}The PageRank values and Harmonic centrality of the social network depicted in Figure~\ref{fig:towns}. Agents are referred as $x_i$, where $x$ denotes the first letter of the towns' color (\textbf{r}ed, \textbf{o}range, \textbf{b}lue, \textbf{g}reen, \textbf{p}ink,) and $i$ denotes the agents location in the town (\textbf{t}op, \textbf{m}iddle, \textbf{b}ottom). Bold numbers represent the maximum value among towns.}
\begin{tabular}{llllll}
Agent/ & PageRank       & Harmonic   & Agent/      & PageRank       & Harmonic   \\
town   & $(\alpha=0.8)$ & centrality & town        & $(\alpha=0.8)$ & centrality \\ \hline
$r_t$  & 0.0770         & 0.571429   & $g_m$       & 0.1259         & 0.214286   \\
$r_m$  & 0.0770         & 0.571429   & $g_b$       & 0.0660         & 0.52381    \\
$r_b$  & 0.0770         & 0.571429   & $p_t$       & 0.0469         & 0.142857   \\
$o_t$  & 0.0547         & 0.488095   & $p_m$       & 0.0469         & 0.142857   \\
$o_m$  & 0.0653         & 0.535714   & $p_b$       & 0.0469         & 0.142857   \\ \cline{4-6}
$o_b$  & 0.0653         & 0.535714   & Red town (avg)   & 0.0770         & $\mathbf{0.5714}$     \\
$b_t$  & 0.0653         & 0.535714   & Orange town (avg)& 0.0617         & 0.5198     \\
$b_m$  & 0.0653         & 0.535714   & Blue town   (avg)& 0.0617         & 0.5198     \\
$b_b$  & 0.0547         & 0.488095   & Green town (avg) & $\mathbf{0.0859}$         & 0.4206     \\
$g_t$  & 0.0660         & 0.52381    & Pink town   (avg)& 0.0469         & 0.1428
\end{tabular}
\end{table}

PageRank correctly predicts the spreading potential of the towns, in the sense, that larger PageRank values correspond to greater spreading capabilities. Harmonic centrality is far less effective, as it ranks the town with the best spreading potential as the second worst.

From this small example we can deduce a couple of interesting points.

\begin{itemize}
    \item Influence maximization techniques are ineffective in comparing sets of nodes, because the most influential agents might be spread over different sets.
    \item Although some sets contain less influential agents (top and bottom green agents are less influential than any of the red ones), as a group they might be more effective than any other.
    \item Aggregating network centralities at a set level can be a good predictor of the sets' spreading potential.
    \item Some centrality measures are more effective than others. Their performance might also vary depending on the underlying graph structure and the applied diffusion model.
\end{itemize}

To avoid inferring too much from a small example, we demonstrate the above points on a real life social network. Furthermore, we will also provide a method to rank centrality measures by their capability in assessing the spreading potential of node sets.

Lastly, we note that the best way to uncover the spreading potential of a set is by simulation. However, this might be prohibitive for various reasons. Running diffusion simulations are costly, especially since we have to do it many times -- often on a very large network too -- to get a reliable estimate of the set's spreading potential.

But the real difficulty comes from the cardinality of the sets rather than the running time of one simulation. If a marketing campaign consists of a series of events, each tied to a different location, the number of sets we need to assess can grow excessively. For instance, if we have to pick five towns among 100, that already means more than 75 million combinations that each need an evaluation by simulation. Calculating influence proxy measures, that is, network centralities that were specifically designed with the underlying problem in mind, is a much cheaper and sometimes the only feasible solution. In our example, the analyst would select the five towns with the highest average node centrality.

\section{Methodology}

In this section, -- following a brief overview of the SRD method -- we introduce our proposed testing framework.

\subsection{SRD}

The input of an SRD analysis is an $n \times m$ matrix where the first $m-1$ columns represent the methods we would like to compare, while the rows represent the measured properties. In our context, columns represent centrality measures (influence proxies) and the rows are the samples we have taken from the node set (potential target groups, \emph{e.g.}\ towns). The last column of the matrix has a special role. It contains the benchmark values, called \emph{references}, which form the basis of comparison. From the input matrix we compose a \emph{ranking matrix} by replacing each value in a column -- in order of magnitude -- by its rank. Ties are resolved by fractional ranking, \emph{i.e.}\ by giving each tied value the arithmetic average of their ranks. SRD values are obtained by computing the city block (Manhattan) distances between the column ranks and the reference ranking. In our paper, the reference values are given externally (they are the average spread of the sample sets in the simulation), but in some applications, the reference values are extracted from the first $m-1$ columns, this process is called \emph{data fusion} \citep{Willett2013}. Depending on the type of data, this can be done by a number of ways (taking the average, median or minimum/maximum). Since we have an external benchmark, we don't delve into the intricacies of data fusion. A more detailed explanation with a numerical example can be found in \citep{Sziklai2020}.

SRD values are normalised by the maximum possible distance between two rankings of size $n$. In this way, problems of different sizes can be compared. A small SRD score indicates that the solution is close to the reference (ranks the rows almost or entirely in the same order as the latter). The differences in SRD values induce a ranking between the solutions. SRD calculation is followed by two validation steps.

\begin{description}
  \item[i.] The permutation test (sometimes called randomisation test, denoted by CRRN = comparison of ranks with random numbers) shows whether the rankings are comparable with a ranking taken at random. SRD values follow a discrete distribution that depends on the number of rows and the presence of ties. If $n$ exceeds 13 and there are no ties, the distribution can be approximated with the normal distribution
      . By convention, we accept those solutions that are below 0.05, that is, below the 5\% significance threshold. Between 5-95\% solutions are not distinguishable from random ranking, while above 95\% the solution seems to rank the objects in a reverse order (with 5\% significance).
  \item[ii.] The second validation option is called cross-validation (CV) and assigns uncertainties to the SRD values. CV enables us to statistically test wether two solutions are essentially the same or not, that is, whether the induced column rankings come from the same distribution.  We assume that the rows are independent in the sense that removing some rows from the matrix does not change the values in the remaining cells. CV proceeds by taking $\ell$ samples from the rows and computing the SRD values for each subset. Different CV techniques (Wilcoxon, Dietterich, Alpaydin) sample the rows in different ways \citep{Sziklai2021b}. Here we opted for 8-fold CV coupled with the Wilcoxon matched pair signed rank test (henceforward Wilcoxon test). For small row sizes $(n\le7)$ leave-one-out CV is applied, each row is left out once. For larger row sizes (as in this paper) a reasonable choice is leaving out $\lceil n/\ell \rceil$ rows uniformly and randomly in each fold. The obtained SRD values are then compared with the Wilcoxon test with the usual significance level of 0.05 for each pair of solutions.
\end{description}

\hfill\\
\textbf{Example}
\hfill\\

Table \ref{tab:toy_example} features a toy example. Suppose we would like to compare two solutions along five properties for which we have reference values. The input table shows how the two solutions perform according to the different properties. The ranking matrix converts these values to ranks. Note that there is a three-way tie for Solution 2. Since these are the 2nd, 3rd and 4th largest values in that column they all get the average rank of 3. Similarly, in the reference column the 3rd and 4th largest value coincide, thus they get an average rank of 3.5.

We compute the SRD scores by taking the absolute differences between the solution's and the reference's ranks row-wise and summing them up. For the first solution, this is calculated as follows:

$$6=|1-2|+|2-1|+|3-5|+|5-3.5|+|4-3.5|.$$

The maximum distance between two rankings that rank objects from 1 to 5 is 12. Thus, we divided the SRD scores by 12 to obtain the normalized SRD values. In the permutation test we compare these values to the 5\% thresholds of the discrete distribution that SRD follows. For $n=5$ we accept those solutions (with 5\% significance) which have a normalized score of $0.25$ or less\footnote{This value comes from the discrete SRD distribution which depends on the number of rows and the presence of ties.}. Compared to the reference Solution 1 seems to rank the properties randomly, while Solution 2 passes the test.

%

\renewcommand{\arraystretch}{1.1}

\begin{table}
\caption{\textbf{\label{tab:toy_example}SRD computation} nSRD stands for normalized SRD.}
\begin{tabular}{lllllllll}
                           & \multicolumn{3}{c}{\cellcolor[HTML]{EFEFEF}Input data} &  &                            & \multicolumn{3}{c}{\cellcolor[HTML]{EFEFEF}Ranking matrix} \\
\multicolumn{1}{l|}{}      & Solution 1.         & Solution 2.        & Ref.        &  & \multicolumn{1}{l|}{}      & Solution 1.          & Solution 2.          & Ref.         \\ \cline{1-4} \cline{6-9}
\multicolumn{1}{l|}{Prop1} & 0.37                & 0.65               & 0.49        & $\longrightarrow$ & \multicolumn{1}{l|}{Prop1} & 1                    & 3                    & 2            \\
\multicolumn{1}{l|}{Prop2} & 0.51                & 0.14               & 0.34        &  & \multicolumn{1}{l|}{Prop2} & 2                    & 1                    & 1            \\
\multicolumn{1}{l|}{Prop3} & 0.82                & 0.88               & 1           & $\longrightarrow$ & \multicolumn{1}{l|}{Prop3} & 3                    & 5                    & 5            \\
\multicolumn{1}{l|}{Prop4} & 0.93                & 0.65               & 0.84        &  & \multicolumn{1}{l|}{Prop4} & 5                    & 3                    & 3.5          \\
\multicolumn{1}{l|}{Prop5} & 0.88                & 0.65               & 0.84        & $\longrightarrow$ & \multicolumn{1}{l|}{Prop5} & 4                    & 3                    & 3.5          \\ \cline{6-9}
                           &                     &                    &             &  & \multicolumn{1}{l|}{SRD}   & 6                    & 2                    &              \\
                           &                     &                    &             &  & \multicolumn{1}{l|}{nSRD}  & 0.5                  & $0.1\overset{.}{6}$                 &
\end{tabular}
\end{table}

\subsection{Testing framework for the spreading potential problem}

Here we describe step by step, how influence measures can be compared under the spreading potential problem framework by using SRD.

\begin{enumerate}
  \item We determine the influence measures (centralities) for each node of the network.
  \item We take $n$ samples of size $q$ from the node set.
  \item For each set we calculate its average\footnote{Depending on the data, alternative statistical aggregates, \emph{e.g.} the sum (for unequal set sizes) or the median (in the presence of outliers), can be applied.} centrality according to each measure.
  \item We run a Monte Carlo simulation with diffusion model $D$ for each of the sample sets and observe their performance, that is their average spread (in the percentage of nodes) in the simulation.
  \item We compile the ranking matrix from the values obtained in step 3 and step 4. These correspond to the solution columns and the reference column respectively.
  \item We compute the SRD values and validate our results.
\end{enumerate}

To demonstrate why such a framework is needed, we add an external validation step. We calculate the top choices of each measure and observe how they perform under the chosen diffusion model. We compare the latter with the ranking obtained from the SRD test. By proving that these two rankings disagree, we can conclude that IM and SP are two separate problems and statistical testing is essential for uncovering the predictive power of influence maximization proxies.

Parameters, $n$ and $q$ should be chosen in accordance with the diffusion model and the size of the network. The network we used for the simulation has 271 913 nodes with 2 712 587 undirected edges (friendships). We converted the network to a directed graph by doubling the edges. Two factors influenced our choice of $n$ and $q$. Firstly, we had to keep the size relatively low to keep the running time at a reasonable level. Secondly, the data does not vary enough if the size of the set is too small or too large, as either nobody or everybody would be infected beyond the sample set. Thus we set $(q=500)$. The number of sample sets ($n=21$) was chosen to be large enough for the SRD scores to be approximately normally distributed and the cross-validation was easier to carry out this way. Further increasing $n$ does not seem to enhance the power of the test, while it does add to the running time. Both for the SRD calculation and for the external validation we ran 5000 simulations of the corresponding sets.

In general, these parameters can be tailored to the needs of the tester, considering the underlying network, diffusion model and available computational capacity.

\section{Diffusion models}\label{sec:diff_models}

A central observation of diffusion theory is that the line that separates a failed cascade from a successful one is very thin. \citet{Granovetter1978} was the first to offer an explanation for this phenomenon. He assumed that each agent has a threshold value, and an agent becomes an adopter only if the amount of influence it receives exceeds the threshold. He uses an illuminating example of how a peaceful protest becomes a riot. A violent action triggers other agents who in turn also become agitated, and a domino effect ensues. However, much depends on the threshold distribution. Despite the similarities, network diffusion can substantially differ depending on the type of contagion we deal with. A basic difference stems from the possible adopting behaviors. In most models, agents can adopt two or three distinct states: susceptible, infected and recovered. Consequently, the literature distinguishes between SI, SIS and SIR types of models.

In the conventional IM setup a seed set of users are fixed as adopters (or are infected), then the results of diffusion is observed. In the SI model, agents who become adopters stay so until the end of simulation \citep{Carnes2007,Chen2009,Kim2013}. A real life example of such a diffusion can be a profound innovation that eventually conquers the whole network (e.g.\ mobile phones, internet). In SIS models, agents can change from susceptible to adopter and then back to susceptible again \citep{Kitsak2010,Barbillon2020}. For example, an innovation or rumour that can die out behaves this way. Another example would be of a service that agents can unsubscribe from and subscribe again if they want. In SIR models, agents can get 'cured' and switch from adopter to recovered status \citep{Yang2007,Gong2018,Bucur2020}. Recovered agents enjoy temporary or lasting immunity to addictive misbehavior (\emph{e.g.}\ to online gaming). Other models allow users to self-activate themselves through spontaneous user adoption \citep{Sun2020}.

In this paper, we feature a classical SI type: the Linear Threshold (LT) model. We represent our network $G$ with a graph $(V,E)$, where $V$ is the set of nodes and $E$ is the set of arcs (directed edges) between the nodes. In LT, each node, $\mathbf{v}$ chooses a threshold value $t_{\mathbf{v}}$ uniform at random from the interval $[0,1]$. Similarly, each edge $e$ is assigned a weight $w_e$  from $[0,1]$, such that, for each node the sum of weights of the entering arcs add up to 1, formally

$$\sum_{\{e\in E| e=(\mathbf{u},\mathbf{v}), \mathbf{u}\in V\}}w_e=1. \quad\quad \mbox{ for all } \mathbf{v}\in V.$$

One round of simulation for a sample set $S_i$ took the following steps.

\begin{enumerate}
  \item Generate random node thresholds and arc weights for each $\mathbf{v}\in V$ and $e \in E$.
  \item Activate each node in $V$ that belongs to $S_i$.
  \item Mark each leaving arc of the active nodes.
  \item For each $\mathbf{u}\in V\setminus S_i$ sum up the weights of the marked entering arcs. If the sum exceeds the node's threshold activate the node.
  \item If there were new activations go back to step 3.
  \item Output the spread \emph{i.e.}\ the percentage of active nodes.
\end{enumerate}

We ran 5000 simulations and took the average of spread values. We can think of the result as the expected percentage of individuals that the campaign will reach if $S_i$ is chosen as initial spreaders.

Overall, diffusion depends on how risk-averse agents are and how much influence they bear on one another. The former is parameterised by the node thresholds, the latter by the arc weights. Since we have each agent's day of registration, we could derive these values for our real-life network. We choose to randomise these values instead. This enables us to run a Monte Carlo simulation and increase the reliability of our analysis.

\section{Influence maximization proxies}\label{sec:IF_proxies}

In this section, we give a brief overview of the measures we employed in our analysis.
\\\hfill\\
\noindent Among the classical centrality measures, we included \textbf{degree} and \textbf{Harmonic centrality}. The former is a self-explanatory benchmark and the latter is a distance-based measure proposed by \citet{Marchiori2000}. Harmonic centrality of a node, $\mathbf{v}$ is the sum the reciprocal of distances between $\mathbf{v}$ and every other node in the network. For disconnected node pairs, the distance is infinite, thus the reciprocal is defined as zero. A peripheral node lies far away from most of the nodes. Thus, the reciprocal of the distances will be small which yields a small centrality value.
\\\hfill\\
\textbf{PageRank}, introduced by \citet{Page99}, is a spectral measure a close relative of \emph{Eigenvector centrality} \citep{Bonacich1972}. Eigenvector centrality may lead to misleading valuations if the underlying graph is not strongly connected. PageRank overcomes this difficulty by (i) connecting sink nodes (\emph{i.e.}\ nodes with no leaving arc) with every other node through a link and (ii) redistributing some value uniformly among the nodes. The latter is parameterised by the so called damping factor, $\alpha\in(0,1)$. The method is best described as a stochastic process. Suppose we start a random walk from an arbitrary node of the network. If anytime we hit a sink node, we restart the walk by choosing a node uniform at random from the node set. After each step, we have a $(1-\alpha)$ probability to teleport to a random node. The probability that we occupy node $\mathbf{v}$ as the number of steps tends to infinity is the PageRank value of node $\mathbf{v}$. The idea was to model random surfing on the World Wide Web. PageRank is a core element of Google's search engine, but the algorithm is used in a wide variety of applications. The damping value is most commonly chosen from the interval $(0.7,0.9)$, here we opted for $\alpha=0.8$.
\\\hfill\\
\textbf{LeaderRank} is a version of PageRank that gets rid of the inconvenient damping parameter by introducing a so called \emph{ground node} \citep{Lu2011}.\footnote{This idea was also entertained by \cite{Tomlin2003}} The ground node is connected to every other node by a bidirectional link which makes the network strongly connected. There is no need to restart a random walk, as there are no sink nodes or inescapable components.
\\\hfill\\
\textbf{Generalised Degree Discount} (GDD) introduced by \citet{Wang2016} is a suggested improvement on \emph{Degree Discount} \citep{Chen2009} which was developed specifically for the \emph{independent cascade} model. The latter is an SI diffusion model where each active node has a single chance to infect its neighbours, transmission occurring with the probability specified by the arc weights. Discount Degree constructs a spreader group of size $q$ starting from the empty set and adding nodes one by one using a simple heuristic. It primarily looks at the degree of the nodes but also takes into account how many of their neighbours are spreaders. GDD takes this idea one step further and also considers how many of the neighbours' neighbours are spreaders. The spreading parameter of the algorithm was chosen to be $0.05$.
\\\hfill\\
\textbf{$k$-core}, sometimes referred to as, $k$-shell exposes the onion-like structure of the network \citep{Seidman1983,Kitsak2010}. First, it successively removes nodes with only one neighbours from the graph. These are assigned a $k$-core value of 1. Then it removes nodes with two or less neighbours and labels them with a $k$-core value of 2. The process is continued in the same manner until every node is classified. In this way, every node of a path or a star graph is assigned a $k$-core value of 1, while nodes of a cycle will have a $k$-core value of 2.
\\\hfill\\
\textbf{Linear Threshold Centrality} (LTC) was, as the name suggests, designed for the Linear Threshold model \citep{Riquelme2018}. Given a network, $G$ with node thresholds and arc weights, LTC of a node $\mathbf{v}$ represents the fraction of nodes that $\mathbf{v}$ and its neighbours would infect under the Linear Threshold model. In the simulation, we derived the node and arc weights that is needed for the LTC calculation from a simple heuristics: each arc's weight is defined as 1 and node thresholds was defined as $0.7$ times the node degree\footnote{We ran LTC and GDD under various parameters and chose the one which performed the best during the external validation.}.
\\\hfill\\
This is by no means a comprehensive list, there are other proxy and centrality measures, \emph{e.g.}\ HITS \citep{Kleinberg99}, DegreeDistance \citep{Sheikhahmadi2015}, IRIE \citep{Jung2012}, GroupPR \citep{Liu2014}, Shapley-value based centrality \citep{Michalak2013}, Top Candidate method \citep{Sziklai2021a}, for more see the survey of \citet{Li2018}. Our goal is to show how such measures should be compared with each other. Announcing a clear winner falls outside the scope of this manuscript. A real ranking analysis should always consider the diffusion model and a network characteristics carefully, as different algorithms will thrive in different environments.

\section{Ranking analysis}

Table \ref{tab:raw} compiles how the 21 sample sets perform according to various influence measures. The last column shows the percentage of nodes that the samples managed to activate on average in the diffusion simulation. The ranking matrix together with the SRD values are featured in Table~\ref{tab:rank}.

\begin{table}
\caption{\label{tab:raw}\textbf{Avg. node centrality.} Abbreviations: PR -- PageRank, LR -- LeaderRank, Harm -- Harmonic centrality, GDD005 -- Generalized Degree Discount with $\alpha=0.05$, LTC07 -- Linear Threshold Centrality with $\alpha=0.7$. Avg. spread denotes the percentage of nodes the sample sets managed to infect on average over 5000 runs.  }\scriptsize
\begin{tabular}{c|cccccccc}
Sample    & \multirow{2}{*}{PR}    & \multirow{2}{*}{$k$-core} & \multirow{2}{*}{LR}    & \multirow{2}{*}{Harm}    & \multirow{2}{*}{GDD005} & \multirow{2}{*}{LTC07}  & \multirow{2}{*}{Degree}   & Avg.   \\
       ID &       &         &       &         &        &        &          &  spread \\
\hline
S1        & 3.623 & 10.152  & 3.631 & 0.21970 & 32.924 & 22.522 & 19.686 & 3.271\%     \\
S2        & 3.664 & 10.532  & 3.707 & 0.22095 & 33.660 & 22.964 & 20.132 & 3.351\%     \\
S3        & 3.620 & 10.248  & 3.601 & 0.22047 & 32.976 & 22.362 & 19.500 & 3.275\%     \\
S4        & 3.698 & 10.078  & 3.695 & 0.21974 & 32.947 & 23.014 & 20.058 & 3.337\%     \\
S5        & 3.663 & 10.226  & 3.627 & 0.21987 & 33.037 & 22.570 & 19.664 & 3.294\%     \\
S6        & 3.462 & 10.100  & 3.402 & 0.21928 & 32.580 & 20.970 & 18.300 & 3.074\%     \\
S7        & 3.790 & 10.606  & 3.846 & 0.22122 & 34.108 & 23.932 & 20.972 & 3.461\%     \\
S8        & 3.802 & 10.838  & 3.825 & 0.22153 & 35.055 & 23.942 & 20.848 & 3.442\%     \\
S9        & 3.525 & 10.036  & 3.510 & 0.21949 & 32.467 & 21.768 & 18.948 & 3.162\%     \\
S10       & 3.628 & 10.560  & 3.609 & 0.22096 & 33.875 & 22.334 & 19.538 & 3.265\%     \\
S11       & 3.594 & 10.334  & 3.598 & 0.21963 & 33.431 & 22.254 & 19.486 & 3.239\%     \\
S12       & 3.713 & 10.498  & 3.735 & 0.22128 & 33.744 & 23.136 & 20.284 & 3.379\%     \\
S13       & 3.737 & 10.204  & 3.710 & 0.21952 & 33.069 & 23.148 & 20.166 & 3.355\%     \\
S14       & 3.712 & 10.058  & 3.696 & 0.21932 & 33.288 & 23.000 & 20.076 & 3.333\%     \\
S15       & 3.723 & 10.500  & 3.728 & 0.22154 & 33.826 & 23.184 & 20.268 & 3.367\%     \\
S16       & 3.826 & 10.786  & 3.794 & 0.22172 & 34.869 & 23.680 & 20.658 & 3.424\%     \\
S17       & 3.782 & 10.446  & 3.813 & 0.22065 & 33.879 & 23.678 & 20.764 & 3.455\%     \\
S18       & 3.762 & 10.602  & 3.775 & 0.22153 & 34.239 & 23.438 & 20.544 & 3.413\%     \\
S19       & 3.635 & 10.382  & 3.628 & 0.22006 & 33.450 & 22.472 & 19.662 & 3.281\%     \\
S20       & 3.577 & 9.964   & 3.530 & 0.21888 & 32.822 & 21.864 & 19.072 & 3.199\%     \\
S21       & 3.652 & 10.672  & 3.665 & 0.22120 & 34.361 & 22.836 & 19.874 & 3.293\%
\end{tabular}
\end{table}

\begin{table}[t]
\caption{\label{tab:rank}\textbf{Ranking matrix.} Data compiled from Table~\ref{tab:raw}. Spreads closer than 0.005\% to each other were considered as a tie. nSRD stands for normalised SRD.}\scriptsize
\begin{tabular}{c|cccccccc}
Sample    & \multirow{2}{*}{PR}    & \multirow{2}{*}{$k$-core} & \multirow{2}{*}{LR}    & \multirow{2}{*}{Harm}    & \multirow{2}{*}{GDD005} & \multirow{2}{*}{LTC07}  & \multirow{2}{*}{Degree}   & Reference   \\
       ID &       &         &       &         &        &        &          & ranking  \\
\hline
S1  & 6& 6& 9& 7& 4& 8   & 9& 6.5   \\
S2  & 11& 15   & 13   & 13 & 12 & 11  & 13 & 13.5  \\
S3  & 5& 9& 5& 11 & 6& 6   & 5& 6.5   \\
S4  & 12& 4& 11   & 8& 5& 13  & 11 & 11.5  \\
S5  & 10& 8& 7& 9& 7& 9   & 8& 9.5   \\
S6  & 1& 5& 1& 2& 2& 1   & 1& 1 \\
S7  & 19& 18   & 21   & 16 & 17 & 20  & 21 & 21\\
S8  & 20& 21   & 20   & 19 & 21 & 21  & 20 & 19\\
S9  & 2& 2& 2& 4& 1& 2   & 2& 2 \\
S10 & 7& 16   & 6& 14 & 15 & 5   & 6& 5 \\
S11 & 4& 10   & 4& 6& 10 & 4   & 4& 4 \\
S12 & 14& 13   & 16   & 17 & 13 & 14  & 16 & 16\\
S13 & 16& 7& 14   & 5& 8& 15  & 14 & 13.5  \\
S14 & 13& 3& 12   & 3& 9& 12  & 12 & 11.5  \\
S15 & 15& 14   & 15   & 20 & 14 & 16  & 15 & 15\\
S16 & 21& 20   & 18   & 21 & 20 & 19  & 18 & 18\\
S17 & 18& 12   & 19   & 12 & 16 & 18  & 19 & 20\\
S18 & 17& 17   & 17   & 18 & 18 & 17  & 17 & 17\\
S19 & 8& 11   & 8& 10 & 11 & 7   & 7& 8 \\
S20 & 3& 1& 3& 1& 3& 3   & 3& 3 \\
S21 & 9& 19   & 10   & 15 & 19 & 10  & 10 & 9.5 \\
\hline
SRD & 22    & 83      & 12   & 73     & 69    & 19    & 12    & 0          \\
nSRD& 0.100  & 0.377    & 0.054 & 0.336   & 0.313  & 0.086  & 0.054  & 0
\end{tabular}
\end{table}

\begin{figure}[t]
  \centering
  \includegraphics[width=12cm]{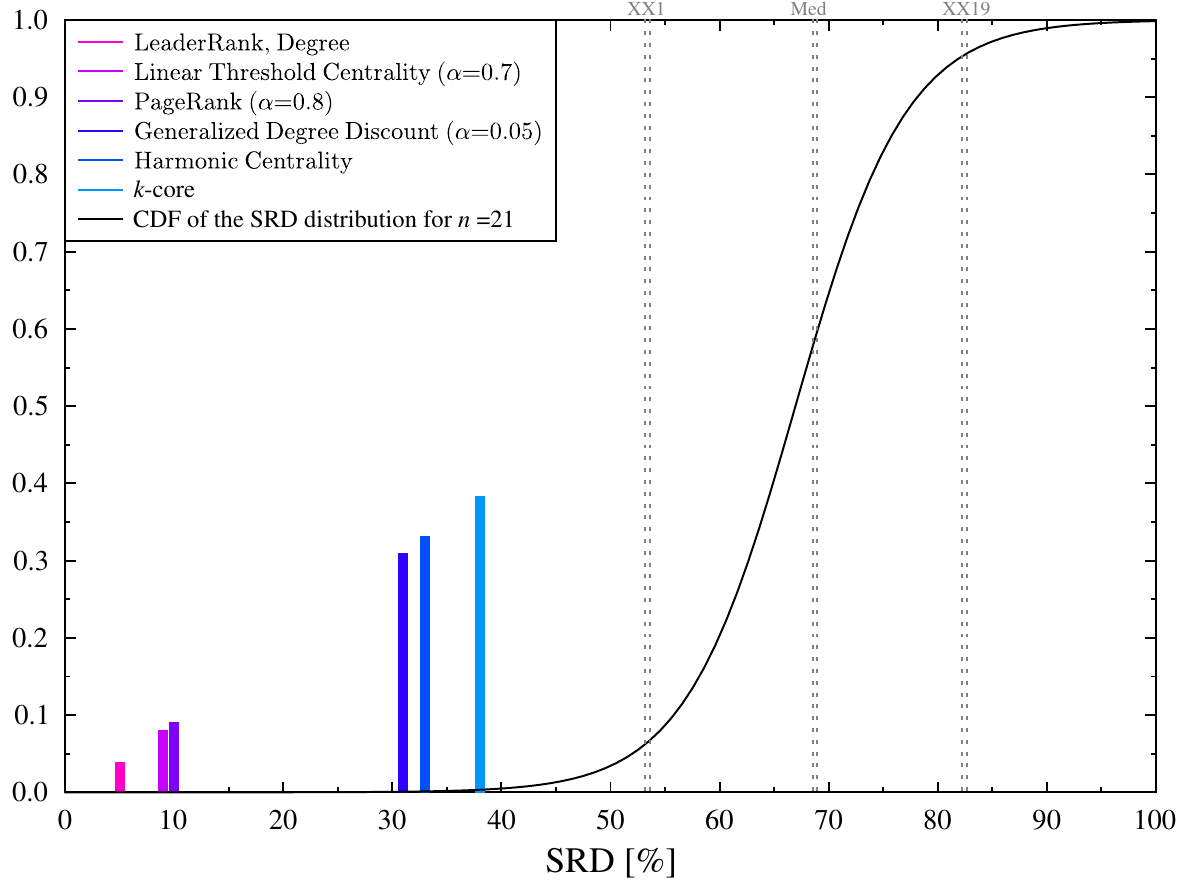}\\
  \caption{\textbf{CRRN test.}  For sake of visualization, the colored bars' height is equal to their normalized SRD values and they follow the same order as in the legend. The black curve is a continuous approximation of the cumulative distribution function of the random SRD values. All (normalised) SRD values fall outside the 5\% threshold (XX1: 5\% threshold, Med: Median, XX19: 95\% threshold).}\label{fig:CRRN}
\end{figure}

All of the methods fall outside the 5\% threshold, that is, they rank the sample sets more or less correctly (on Fig.~\ref{fig:CRRN} they are to the left of the XX1 line). LeaderRank and Degree have the best (lowest) SRD score, closely followed by the Linear Threshold Centrality and PageRank. It shouldn't be surprising that LTC performs well under the Linear Threshold model. Also, we shouldn't hold the relatively high SRD score against the Generalised Degree Discount algorithm, as it was primarily designed for the Independent Cascade model. The take-home message is that despite its simplicity, Degree has still a very good predictive power. On the other hand, $k$-core and Harmonic Centrality seem to be the least adequate proxies for this particular environment (running the LT model on a social network). Let us stress again that the scope of the results is limited to the underlying network type and the applied diffusion model.

\begin{figure}[t]
  \centering
  \includegraphics[width=12cm]{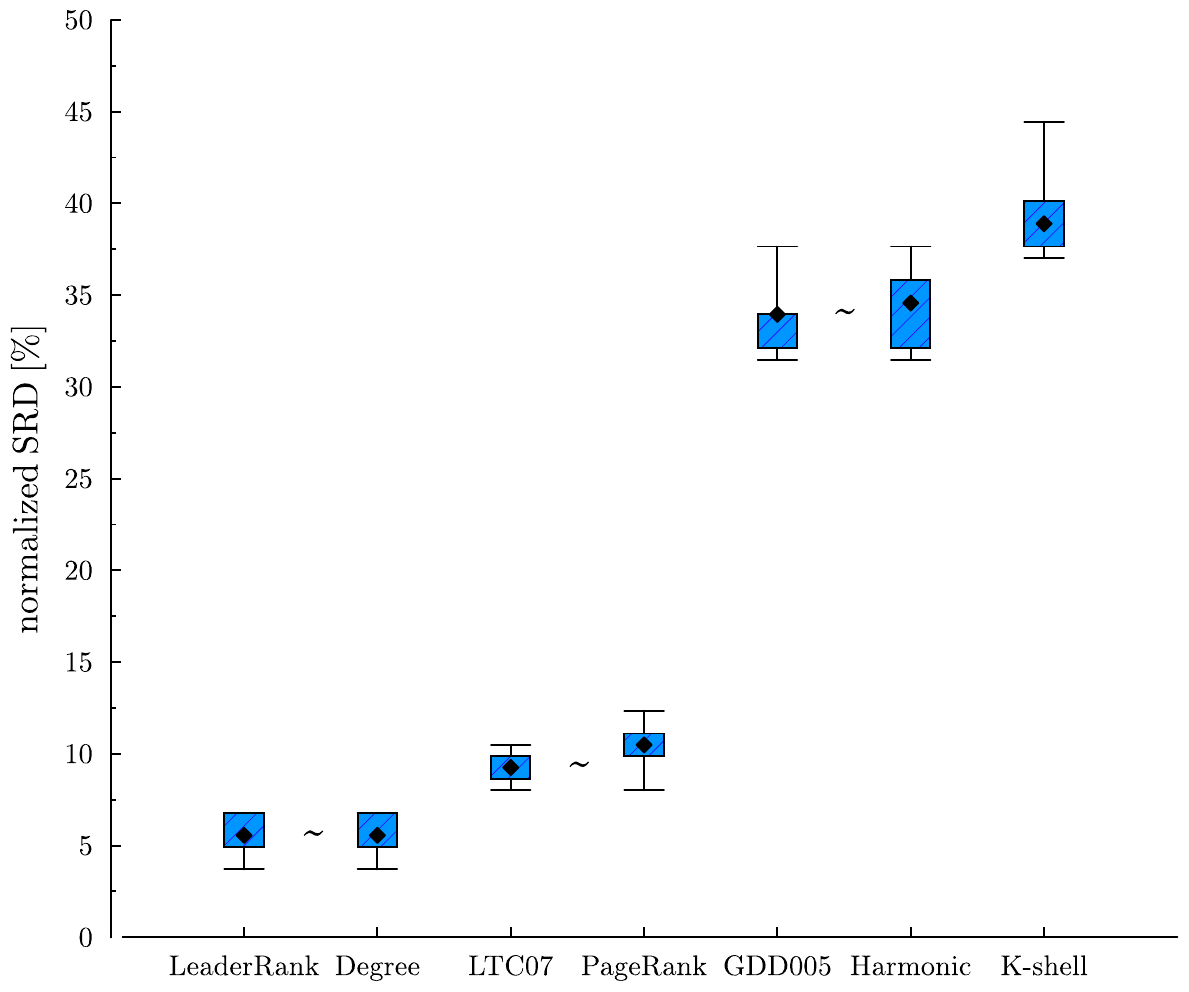}\\
  \caption{\textbf{Cross-validation.} The boxplot shows the median (black diamond), Q1/Q3 (blue box) and min/max values. The measures are ranked from left to right by the median values. The '\texttildelow' sign between two neighbouring measures indicate that we couldn't reject the null hypothesis, the Wilcoxon test found no significant difference between those rankings.}\label{fig:CV}
\end{figure}

Cross-validation reveals how the methods are grouped. According to the Wilcoxon test, there is no significant difference between LeaderRank and degree. Similarly, the rankings induced by LTC and PageRank seem to come from the same distribution, and the same goes for GDD and Harmonic centrality. Fig.~\ref{fig:CV} shows how the measures are ranked.

Our goal was not to find the best influence measure for SP, that would need a more comprehensive comparison, but rather to show that SP and IM are two different problems. So let us take a look at how the top choices of these measures perform in the simulation. We took the 500 highest ranked agents for each measure. If the 500th and 501st agents tied with each other, we discarded agents with the same score one by one randomly until the size of the set became 500. We ran 5000 simulations to obtain the average spread of the measures.

\begin{table}
\small
\caption{\label{tab:ext_valid}\textbf{Performance of the top choices of the measures (IM) on a simulation of size 5000 compared to the ranking obtained by the SRD test (SP). }In the IM ranking, PageRank's and LTC's performance is very close, so we may consider it as a tie. If we take into account the result of the Wilcoxon test there are ties in the SP ranking as well (see the ranks in brackets).  }
\begin{tabular}{l|lllllll}
              & PageRank       & $k$-core& LeaderRank   & Harmonic       & GDD005        & LTC07         & Degree  \\
               \hline
Avg. spread   & 21.021\%       & 4.559\% & 21.099\%     & 16.243\%       & 21.214\%      & 21.024\%      & 21.157\% \\
Std. error    & 0.019\%        & 0.012\% & 0.020\%      & 0.019\%        & 0.020\%       & 0.020\%       & 0.020\% \\
Rank by IM    & 5 (4.5)        & 7       & 3            & 6              & 1             & 4 (4.5)       & 2       \\
Rank by SP    & 4 (3.5)        & 7       & 1.5          & 6 (5.5)        & 5 (5.5)       & 3 (3.5)       & 1.5
\end{tabular}

\end{table}

Table~\ref{tab:ext_valid} displays the results. There is a substantial difference between the ranking induced by the performance of the top choices (IM) and the ranking resulting from the SRD analysis (SP), especially if we consider the results of the Wilcoxon test as well. GDD, which was among the poor performers, becomes the very best. On the other hand, LeaderRank which shared the first place with Degree, is now on the 3rd place. The avg.\ spread of PageRank and LTC07 are very close, thus we display their ranks in two ways (the tied ranks are shown in brackets).

Comparing the top choices of these measures only reveals which of them identifies the best spreaders. However, this basic test does not let us know the algorithms' real predictive power for an arbitrary set of agents.

\section{Conclusion}

We described a realistic decision situation that occurs in viral marketing campaigns. A company that organises a series of public events wants to maximize social influence by selecting suitable locations/audiences. For this reason the company must evaluate the collective impact of the selected audiences.
As exact computation is infeasible, the potential is  estimated by influence proxies (\emph{e.g.}\ network centralities). The question is which measure suits our needs the best? The spreading potential problem resembles to the influence maximization at first glance. To prove that they are different we proposed a testing framework that is capable to rank influence measures by their predictive power. For this, we used a novel comparative statistical test, the Sum of Ranking Differences.

We demonstrated our results on a real-life social network. We took samples from the node set and created multiple rankings using various centrality measures. We compared these rankings to a reference ranking induced by the spreading potential of these sets in a Linear Threshold diffusion model. The algorithm whose ranking is the closest to the reference has the best predictive power. We compared this ranking to a ranking obtained by running a simulation with the top choices of these measures (in the manner of IM).

The best performing influence maximization algorithm fared relatively poorly on the spreading potential test. The discrepancy between the two rankings implies that we cannot blindly assume that the algorithm that finds the top spreaders in a network is the best in predicting the spreading potential of an arbitrary set of agents. 

There are several directions that suggest themselves for further research. To find which influence measure performs the best in the spreading potential problem under the linear threshold diffusion model a more comprehensive comparison is needed that involves a wide variety of network centralities that are tested on different types of networks. We conjecture that SP and IM ranks centralities differently under other diffusion models, but this is also a question of future research.

\section*{Acknowledgment}
The authors acknowledge financial help received from National Research, Development and Innovation Office grant numbers KH 130502, K 138970, K 128573, K 138945. This research was supported by the Higher Education Institutional Excellence Program 2020 of the Ministry of Innovation and Technology in the framework of the 'Financial and Public Services' research project (TKP2020-IKA-02) at Corvinus University of Budapest. Bal\'azs R.\ Sziklai is the grantee of the J\'anos Bolyai Research Scholarship of the Hungarian Academy of Sciences.

\bibliography{proxySRD_ref}

\end{document}